\documentclass[prb,twocolumn]{revtex4}
\usepackage{graphicx}
\usepackage{graphics}
\begin{document}
\title{Electronic transport in quantum cascade structures}
\author{C. Koeniguer}
\affiliation{Mat\'eriaux et Ph\'enom\`enes Quantiques, Univ. Paris 7, Case 7021, 2 Pl Jussieu, 75251 PARIS, France}
\author{G. Dubois}
\altaffiliation{Present address : Laboratoire Kastler Brossel, Ecole normale sup\'erieure, 24 rue Lhomond, 75005 PARIS, France}
\affiliation{Mat\'eriaux et Ph\'enom\`enes Quantiques, Univ. Paris 7, Case 7021, 2 Pl Jussieu, 75251 PARIS, France}
\author{A. Gomez}
\affiliation{Mat\'eriaux et Ph\'enom\`enes Quantiques, Univ. Paris 7, Case 7021, 2 Pl Jussieu, 75251 PARIS, France}
\author{V. Berger}
\affiliation{Mat\'eriaux et Ph\'enom\`enes Quantiques, Univ. Paris 7, Case 7021, 2 Pl Jussieu, 75251 PARIS, France}

\begin{abstract}

The transport in complex multiple quantum well heterostructures is theoretically described. The model is focused on quantum cascade detectors, which represent an exciting challenge due to the complexity of the structure containing 7 or 8 quantum wells of different widths. Electronic transport can be fully described without any adjustable parameter. Diffusion from one subband to another is calculated with a standard electron-optical phonon hamiltonian, and the electronic transport results from a parallel flow of electrons using all the possible paths through the different subbands. Finally, the resistance of such a complex device is given by a simple expression, with an excellent agreement with experimental results. This relation involves the sum of transitions rates between subbands, from one period of the device to the next one. This relation appears as an Einstein relation adapted to the case of complex multiple quantum structures.
\end{abstract}

\maketitle
\section{Introduction}

Quantum Well Infrared Photodetectors (QWIPs) have become widely used quantum heterostructures for thermal imaging application during the last ten years \cite{schneider}. Thanks to the very recent introduction of Quantum Cascade Detectors (QCDs) \cite{laure1,graf,laure2}, QWIPs exist now in two modes of electronic transport : photoconductive and  photovoltaic. For the first type (the regular QWIP inserted today in focal plane arrays), an external bias is necessary to drift the electrons. They are characterized by a dark current and so they operate at low temperature. The second type of detectors uses a photovoltaic mode : no external bias is applied. No dark current exists in these devices and they are promising for applications at higher temperature, larger wavelengths~\cite{graf} and smaller pixel areas. In all these structures, however, the electronic transport is a very complex issue and the performances of the devices are finally governed by fundamental physics problems.\\
\begin{figure}[ht]
\begin{center}
\includegraphics[width=7cm]{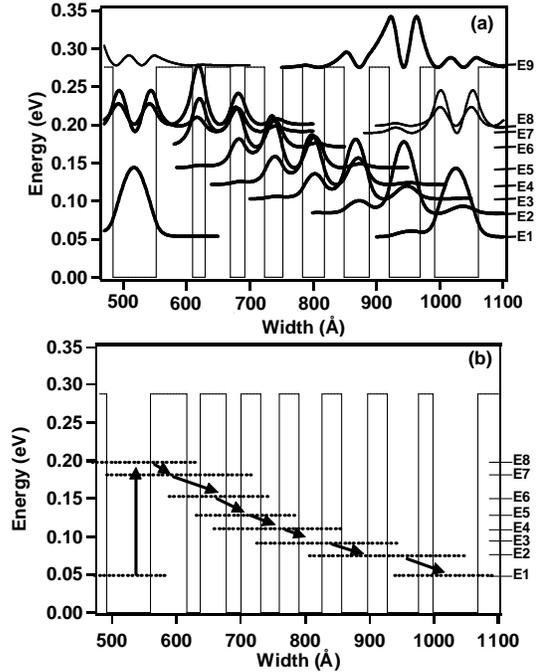}
\newline
\caption{\label{figure1}Presentation of a period of a QCD. 1(a) Conduction band diagram and wave function associated with each energy level of a period. 1(b) Principle of a detection.}
\end{center}
\end{figure}
In a standard QWIP, the transport involves both 2D and 3D electronic states in the quantum well and in the continuum, respectively. The diffusion of electrons from 3D to 2D states (and vice versa) is a particularly difficult theoretical problem\cite{brum,brumbastard}. That is why most models use adjustable parameters such as the capture time \cite{rosencher} or the capture probability, affecting the photoconductive gain \cite{liu}. In quantum cascade lasers, on another side, significant electric fields are applied, resulting in charge transfers. The description of the electronic transport has to involve Schr\"odinger and Poisson equations self-consistantly, which is a source of difficulty as well. With respect to these two devices, the QCD structure is rather simple since it involves only 2D states (and therefore matrix elements can be calculated without difficulty) and no bias is applied : there is no need for Poisson equation, and the flat band condition is an excellent approximation, also because the doping levels are low. For all these reasons, the QCD appears as an archetype of complex quantum heterostructures, a model system for the experimental study of the electronic transport in a complex multiple quantum well structure together with its modeling.  This will be shown in this paper, in which the mechanisms of electronic transport in quantum cascade heterostructures are described. For the first time, electronic transport in such a complex structure is calculated ab initio. By this expression, we mean that the current is calculated without any other adjustable parameters than standard effective masses and band offsets found in the general literature on semiconductors. The details of the QCD structure and the principle of detection are described in Sec.II. In Sec.III, the model describing the electronic transport is presented in details, relying on the calculation of all the transition rates between different subbands due to an electron-optical phonon hamiltonian. Through an analytical derivation of the sum of the different paths that an electron can follow to cross the structure, the resistance of the complex multiple quantum well structure is expressed in a very simple form (Sec. IV). The analogy between this expression and the Einstein relation will be underlined.  

\section{Presentation of a QCD.}
The QCD considered in this paper is a GaAs/AlGaAs heterostructure composed of 40 periods of 7 quantum wells. The quantum wells are made in GaAs. The first quantum well of each period is n-doped in order to populate its first energy level $E_1$ in the conduction band with electrons (the nominal doping concentration is about $5. 10^{11} \text{cm}^{-2}$). Barriers are made in $\text{Al}_{0.34}\text{GaAs}_{0.66}$. The quantum wells (respectively the barriers) have the following widths: 68, 20, 23, 28, 34, 39 and 48 $\mathring{A}$ (respectively 56.5, 39.55, 31, 31, 31, 31, 22.6 $\mathring{A}$). These dimensions have been optimized in order to create the quantum cascade of levels shown in figure 1. Figure 1 (a) presents wave functions associated with each energy subband, in one period of the device. Wave functions associated with each energy subband are delocalized through the heterostructure. Absorption of a photon brings electrons from the first energy level to energy $E_7$ or $E_8$. The first and the second quantum wells are coupled. In this way, electrons are transferred from level $E_7$ or $E_8$ towards energy level $E_6$ mostly through the longitudinal optical phonon interaction. Because of the coupling between the second and the third quantum well, electrons will be relaxed to energy level $E_5$ and so on. Electrons are relaxing from the left to the right of the structure step by step (see figure 1 (b)). Finally the net result under illumination is the transfer of electrons from energy level $E_1$ of one period to energy level $E_1$ of the following period. In this way, the structure acts as a photovoltaic infrared detector.\\

\section{Model of the electronic transport}
As usual with this kind of device, the resistance will be presented in terms of $R_0A$ (where $R_0$ is the resistance of the pixel and $A$ the area of the pixel). Hypotheses about electronic transitions between the different energy levels have to be made for the determination of the $R_0A$ of such a structure. Considering the well and barrier widths, only interactions between electrons and optical phonons (LO-phonons) have been taken into account. The differences between the energy levels are indeed sufficiently high to neglect the influence of the interaction between electrons and acoustical phonons\cite{radovanovic}. All other possible interactions have been neglected too. Interface roughness has low influence on intersubband scattering, although it was possible to measure its influence at very low temperature~\cite{aude} (4 K). Electron-electron interaction is efficient for electrons thermalization inside a subband~\cite{goodnick,harrison}, but negligible for intersubband scattering in our cascade scheme with many subband separations greater than $36 \:$ meV which corresponds to the energy of the optical phonons~\cite{kempa1, kempa2}. The transition rates due to the interaction between electrons and optical phonons will be evaluated first following Ferreira et al. \cite{ferreira}. Starting from an initial state of wavevector $k$ and energy $E$ in the subband $i$, the transition rate $S_{ij}^{a,e}(E)$ towards the subband $j$ (in s\textsuperscript{-1}) is obtained through the integration of a matrix element involving a standard electron-optical-phonon Hamiltonian. This integration involves all the possible final states of energy $E \pm \hbar \omega_{LO}$ in the subband $j$, where $\omega_{LO}$ is the energy of a LO-phonon, the plus or minus sign accounting for absorption or emission of LO-phonons, corresponding to  superscript $a$ or $e$, respectively. This evaluation is made in the parabolic approximation of the energy bands. Transitions rates $S_{ij}^a$ and $S_{ji}^e$ are linked by the following equation :
\begin{equation}
S_{ij}^a(E)=S_{ji}^e(E+\hbar \omega_{LO}\label{eq_sij})
\end{equation}
Finally, the global transition rate $G_{ij}$ between the subband $i$ and subband $j$ is the sum of the two transition rates for absorption of LO-phonons ($G_{ij}^a$), and emission of LO-phonons ($G_{ij}^e$). In order to calculate the global transition rates $G_{ij}^a$ and $G_{ij}^e$, all the initial states of energy $E$ are filled at thermal equilibrium by the Fermi-Dirac occupation factor $f$. Let us first consider the electronic promotion from a subband $i$ to a higher subband $j$ (i.e. $j>i$). The integration on all these states is now performed on the subband $i$:\\
\begin{eqnarray}
G_{ij}^a & = & \int_{\epsilon_j-\hbar \omega_{LO}}^{+\infty} S_{ij}^a(E)f(E)\left(1-f(E+\hbar \omega_{LO})\right)\nonumber \label{eq_gija} \\
&&\times n_{opt} D(E)dE\\
G_{ij}^e & = & \int_{\epsilon_j+\hbar \omega_{LO}}^{+\infty} S_{ij}^e(E)f(E)\left(1-f(E-\hbar \omega_{LO})\right) \nonumber \\
&&\times \left(1+n_{opt}\right) D(E)dE \label{eq_gije}
\end{eqnarray}
$n_{opt}$ is the Bose-Einstein statistic function which accounts for phonon population, $D(E)$ the two-dimensional density of states of the subband $i$ and $\epsilon _j$ is the minimum of energy of the subband $j$. Of course, similar expressions can be written in the case of electronic transfers from a subband $i$ to a lower subband $j$. Figure 2 summarizes the four possible electronic transitions.
\begin{center}
\begin{figure}[ht]
\includegraphics[height=4cm]{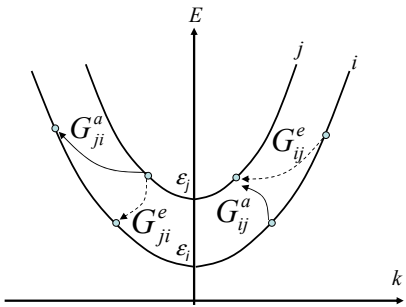}
\caption{\label{figure2}The different types of electronic transitions between two subbands $i$ and $j$}
\end{figure}
\end{center}
At thermodynamical equilibrium, each global coefficient $G_{ij}$ is equal to its reciprocal factor $G_{ji}$. The number of electrons per second transferred from subband $i$ to subband $j$ is equal to the number of electrons per second transferred back from subband $j$ to subband $i$. The net current is null and the population in every subband is stationnary.\\
In the context of a photovoltaic detector, the resistance at $0 \: V$ is a key parameter since it governs the Johnson noise of the device. The calculation of this resistance needs the determination of the current for a very small applied bias. The bias can then be introduced as a perturbation, and therefore we will suppose that the matrix elements and the $S_{ij}(E)$ transitions rates do not depend on the voltage. Indeed, the electronic wavefunctions given by the weakly perturbed potential can be supposed non dependant on the arbitrary small voltage. In the frame of this approximation, the current results only from the variation with the applied bias of the distribution of the population as a function of the energy which affect the $G_{ij}s$ through the occupation factors $f$.\\
As an illustration of the exchanges of electrons between subbands occurring in the structure, Figure 3 represents the main transition rates at thermodynamical equilibrium at 80 K. In addition, these transfer rates are indicated in table I. Transition times $\tau_{ij}$ can be deduced from the global transition rate $G_{ij}$ by:\\
\begin{equation}
\tau _{ij}=n_i/G_{ij}
\end{equation}
where $n_i$ is the two dimensional carrier density associated with the subband $i$. Two consecutive cascades A and B are represented in Figure 3. For sake of clarity, transitions between two levels in the same cascade have been represented separately from transitions between two levels in consecutive cascades. On Figure 3(a), most significant intra-cascade transitions are represented. The typical value of transitions rates between two neighboring levels are between a few $10^{20} \ \text{and} \ 10^{25}\ \text{m}^{-2} \text{s}^{-1}$, corresponding to transition times between a few ps and a few tens of ps for a temperature of 80 K. On Figure 3(b), the inter-cascade transitions are reported. Again, the solid line arrows represent the main transfer rates between one cascade and the following one, but these transitions are now limited to a few $10^{18} \ \text{m}^{-2} \text{s}^{-1}$ at the same temperature (and a corresponding transition time greater than 1 $\mu s$ due also to the low amount of electron promotion to higher subbands for satisfying Fermi Dirac distribution). The dashed line arrows represent other minor transitions.
\begin{figure}[ht]
\begin{center}
\includegraphics[width=8cm]{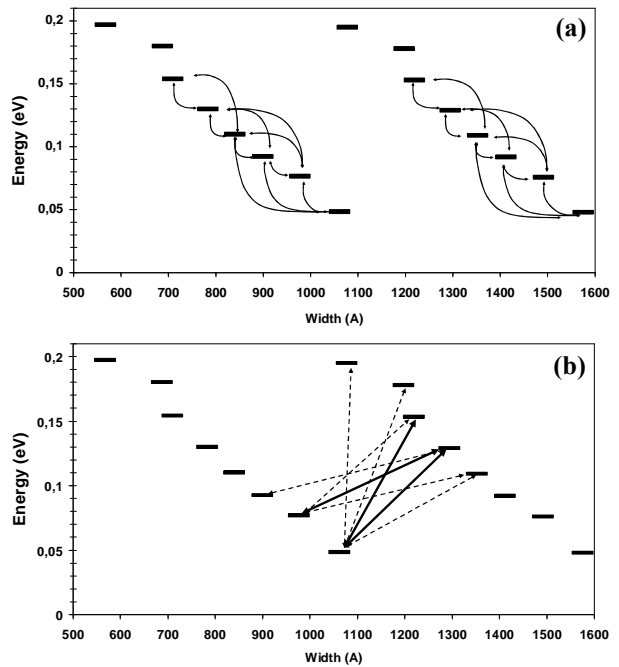}
\caption{\label{figure3}Major transition rates of two consecutive cascades of the real device at 80 K. Figure 3(a) represents the main transition rates inside each cascade (only the transition rates greater than $10^{20}\ \textrm{m}^{-2} \mathrm{s}^{-1}$ are represented). Figure 3(b) represents the main transitions between the cascades : solid lines concern the major transition (greater than $4. \ 10^{18}\ \textrm{m}^{-2} \mathrm{s}^{-1}$) whereas dashed lines represents the other main transitions (greater than $1. \ 10^{18}\ \textrm{m}^{-2} \mathrm{s}^{-1}$ and lower than the major transitions).}
\end{center}
\end{figure}
The comparison between intra- and inter-cascade transitions shows that the electronic mobility is higher inside a cascade than between two consecutive cascades by several orders of magnitude. This justifies our classification between inter- and intra-cascade transitions and has major consequences on the understanding of the transport in the structure.
\begin{table}[!h]
\begin{ruledtabular}
\begin{tabular}{c c}
intra-cascade transition & inter-cascade transition \\
rate ($\text{m}^{-2}\text{s}^{-1}$) & rate ($\text{m}^{-2}\text{s}^{-1}$)\\
\hline
$G_{12}=2.7 \: 10^{25}$&$G_{14}=3.0 \: 10^{18}$\\
$G_{13}=1.0 \: 10^{24}$&$G_{15}=5.1 \: 10^{18}$\\
$G_{14}=1.2 \: 10^{22}$&$G_{16}=5.0 \: 10^{18}$\\
$G_{23}=1.7 \: 10^{24}$&$G_{17}=3.5 \: 10^{18}$\\
$G_{24}=3.8 \: 10^{23}$&$G_{18}=2.3 \: 10^{18}$\\
$G_{25}=1.2 \: 10^{21}$&$G_{24}=1.1 \: 10^{18}$\\
$G_{34}=1.9 \: 10^{23}$&$G_{25}=4.6 \: 10^{18}$\\
$G_{35}=2.2 \: 10^{22}$&$G_{26}=1.3 \: 10^{18}$\\
$G_{45}=1.2 \: 10^{22}$&$G_{35}=1.3 \: 10^{18}$\\
$G_{56}=4.6 \: 10^{20}$&\\
\end{tabular}
\end{ruledtabular}
\caption{\label{tableI}Values of some transition rates in relation to the main electronic transition represented on the figure 3. The left column concern the higher transition rates inside each cascade (intra-cascade transition rate) and the right column gives the higher transition rates between two consecutive cascades (inter-cascade transitions rates). These values are evaluated at 80 K.}
\end{table}
It recalls the situation in a p-n junction, where intra-band thermalization is faster than interband recombination by several orders of magnitude (picoseconds versus nanoseconds). More generally, this is a very general situation in the physics of transport where a system can be divided into two reservoirs separated by a bottleneck for the conduction. The concept of quasi Fermi level can be imported from the p-n junction formalism into our case. Each cascade stays at a quasi thermodynamical equilibrium. Two quasi Fermi levels associated with two consecutive cascades can be considered : the cascade A (resp. B) is associated with the quasi-Fermi energy $E_F^A$ (resp. $E_F^B$). At thermodynamical equilibrium without applying any voltage these two quasi-Fermi levels are equal. The structure is characterized by only one Fermi level. In the presence of an external applied voltage, a gap between the two quasi Fermi levels appears and increases with the voltage : $E_F^A = E_F^B + qV$. Each energy level of the cascade A increases of the same factor $qV$ too. Figure 4 represents the situation of two consecutive cascades under a bias voltage $V$.\\
The main transitions at 80 K between two consecutive cascades occur between subband $E_1^A$ to the subbands $E_5^B$ and $E_6^B$ (transition towards level $E_5^B$ is more efficient than the other transitions). In conclusion, the resistance of a QCD is completely determined at 80 K by a few inter-cascade cross transitions, namely $E_1^A \rightarrow E_5^B$ and $E_1^A \rightarrow E_6^B$ . The optimization of a QCD requires decreasing these transitions rates, thus increasing the resistance and decreasing the noise figure. This is possible by a separation of the wavefunctions, but at the expanse of a lower optical matrix element, and a lower response. All the challenge of the QCD design consists of mastering this trade off.\\
\begin{figure}[ht]
\includegraphics[width=8cm]{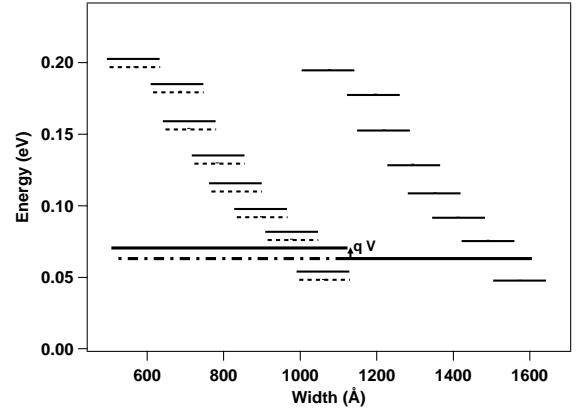}
\caption{\label{figure4}Evolution of the first cascade in relation to the second cascade under a bias voltage. Solid lines represent the position of the energy levels under a bias voltage and dashed lines concern the position of the energy levels without any applied voltage. Fermi levels are represented on this figure in bold lines.}
\end{figure}

\section{Derivation of an einstein relation}

The global current density will be evaluated by counting the electronic transitions between the two consecutive cascades. On one hand, each cascade stays at thermodynamical equilibrium, but on the other hand equilibrium between the two consecutives cascade is broken. As a consequence, transition rates from cascade A to cascade B are not equal to the reciprocal transitions rates from cascade B to cascade A. A global current appears. This global current density is given by:\\
\begin{equation}
J=q\sum_{i\in A}\sum_{j\in B} \left(G_{ij}(V)-G_{ji}(V)\right)
\end{equation}
Let us consider two subbands $i$ and $j$ associated respectively with cascades A and B. The introduction of two quasi Fermi levels implies the differentiation of the two Fermi occupation factors in Eq.~\ref{eq_gija} and Eq.~\ref{eq_gije}. $G_{ij}^a$ and $G_{ji}^e$ can be evaluated by :\\
\begin{eqnarray}
G_{ij}^a(V)&=&\int_{\epsilon_j-\hbar \omega_{LO}}^{+\infty} S_{ij}^a(E)f_A(E)\left(1-f_B(E+\hbar \omega_{LO})\right)\nonumber\\
&&\times n_{opt} D(E)dE\\
G_{ji}^e(V)&=&\int_{\epsilon_j}^{+\infty} S_{ji}^e(E)f_B(E)\left(1-f_A(E-\hbar \omega_{LO})\right)\nonumber \\
&&\times\left(1+n_{opt}\right) D(E)dE
\end{eqnarray}
where $f_A$ and $f_B$ are the Fermi-Dirac occupation factors associated with quasi Fermi level $E_F^A$ and $E_F^B$.
Considering Eq.~\ref{eq_sij}, the difference   is then equal to:
\begin{equation}
G_{ij}^a(V)-G_{ji}^e(V)=\int_{\epsilon_j-\hbar \omega_{LO}}^{+\infty} S_{ij}^a(E) \alpha(E) \left(1-\gamma(E)\right)dE
\end{equation}
where:\\
\begin{eqnarray}
\alpha(E) & = & n_{opt} f_A(E) \left(1-f_B(E+\hbar \omega_{LO})\right)D(E)\\
\gamma(E) & = & \frac{f_B(E+\hbar \omega_{LO})\left(1-f_A(E)\right)\left(1+n_{opt}\right)}{f_A(E)\left(1-f_B(E+\hbar \omega_{LO})\right)n_{opt}}
\end{eqnarray}
Without any applied voltage, the term $\alpha(E)$ is equal to $\alpha^{eq}(E)$ given by:\\
\begin{equation}
\alpha^{eq}(E)=n_{opt} f(E) \left(1-f(E+\hbar \omega_{LO})\right)D(E)
\end{equation}
It corresponds to the first term of the series expansion of $\alpha$ as a function of the voltage $V$. Expressing the Fermi Dirac functions, $\gamma (E)$ is simplified into :\\
\begin{eqnarray}
\gamma(E)&=&\exp\left(\frac{E_F^B-E_F^A}{k_b T} \right)\nonumber \\
&=&\exp\left( \frac{-qV}{k_b T}\right)
\end{eqnarray}
where $T$ is the temperature of the sample and $k_b$ the Boltzman constant.\\
We recall that in the context of infrared photovoltaic detection, applied biases are very small. The Johnson noise is related to the resistance at $0 V$: $R_0$. In this calculation, it is then justified to linearize :\\
\begin{equation}
1-\gamma(E)\approx \frac{q}{k_b T}V
\end{equation}
This leads directly to the linear $I(V)$ behavior of the structure at low bias, through the multiplication by the constant $\alpha ^{eq}$ (calculated with no applied voltage). For little variations of the voltage, the difference   can be approximated by the following equation:\\
\begin{eqnarray}
G_{ij}^a(V)-G_{ji}^e(V) & \approx & \int_{\epsilon_j-\hbar \omega_{LO}}^{+\infty} S_{ij}^a(E) \alpha^{eq} \frac{qV}{k_b T}dE\\
& \approx & G_{ij}^a(V=0V) \frac{qV}{k_b T}
\end{eqnarray}
A similar expression can be found for the difference  :\\
\begin{equation}
G_{ij}^e(V)-G_{ji}^a(V) \approx G_{ij}^e(V=0V) \frac{qV}{k_b T}
\end{equation}
Finally, global current density is so evaluated by the formula:\\
\begin{equation}
J=q\sum_{i\in A}\sum_{j\in B} G_{ij}\frac{q V}{k_b T}
\end{equation}
where the term  $G_{ij}$ is defined by the sum of $G_{ij}^a$ and $G_{ij}^e$ calculated without any applied voltage.
$R_0 A$ can be finally deduced from the last equation:\\
\begin{equation}
R_0 A=\frac{k_b T}{q^2 \sum_{i\in A}\sum_{j\in B} G_{ij}} \label{eq_roa}
\end{equation}

We have previously made the hypothesis that the electronic wavefunctions and the matrix elements are not dependant on the small bias. The current has been calculated as a result from a variation of the distribution of carriers as a function of the energy from cascade A to cascade B. This means that the current appears as a diffusion current, in a system with a non-homogenous chemical potential. In this context, it is expected to find an expression of the $R_0A$ which looks like an Einstein relation. By an identification between $R_0A$ and $l^2/\left(q n_{2D}\mu \right)$ (where $l$ is the period of the structure, $n_{2D}$ is the 2D-density and $\mu$ is the electron mobility) we find indeed again $D / \mu = k_b T / q$, with the diffusion coefficient $D$ expressed classically as a function of the transfer rates at equilibrium $G_{ij}$ by :\\
\begin{equation}
D=\frac{l^2\sum_{i\in A}\sum_{j\in B} G_{ij}}{n_{2D}}
\end{equation}

\begin{figure}[ht]
\includegraphics[width=8cm]{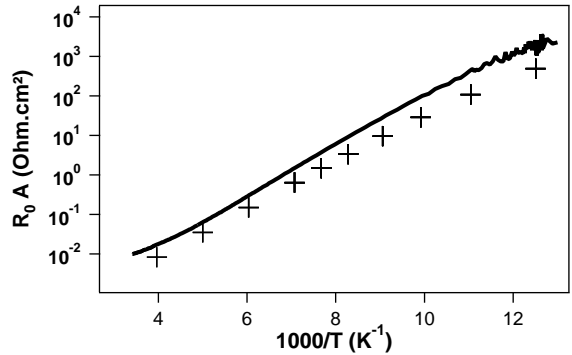}
\caption{\label{figure5}$R_0 A$ as a function of 1000/T, where T is the temperature of the sample. Solid line corresponds to the experimental curve and the markers are the results of the presented model.}
\end{figure}
We find again $D=l^2/\tau$ used in the context of diffusion currents in p-n junctions \cite{rosenchervinter}. It is remarkable in Eq. ~\ref{eq_roa} that the $R_0A$ can be calculated in a structure with no applied bias, considering only the transition rates $G_{ij}$ at equilibrium, from one cascade to the next. Eq.~\ref{eq_roa} appears as the main result of this paper. We have shown that the electronic transport in a complex structure as a QCD can be simplified. The transport is entirely dominated by 2 or 3 matrix elements present in the sum in Eq.~\ref{eq_roa}, which is interpreted as an Einstein relation adapted to QCDs.\\
Figure 5 presents a typical example chosen to illustrate the quality of the modeling : the experimental $R_0A$ of the device as a function of $1000/T$, where $T$ is the temperature of the sample, and the result of our modeling using the previous Eq. ~\ref{eq_roa}. Activation energy deduced from experimental curve lies around $120$ meV. This energy corresponds to the transition between the first energy level $E_1$ and the energy level $E_6$. Dark current originates essentially from electrons transferred from level $E_1$ to level $E_6$, which is in a good agreement with the evaluation of the transfer rates presented in this article. The theoretical modeling provides a good fit of the experimental curve : activation energy deduced from the model lies around $120$ meV too.\\
The resistivity given by the model is slightly lower than the experimental value. We atribute this small discrepancy to a doping level slightly lower than expected. Indeed, including a lower doping level (around $3.\:10^{11}\:\mathrm{cm}^{-2}$) in the model provides a perfect fit of the experimental result, and on the other hand the measurement of the integrated absoprtion agrees very well with a lower value of the doping level with respect to the nominal value. Other small discrepancies between the experimental results and our model could come from other deviations of the structure parameters from the nominal values. The influence of the variations of structure parameters on the resistivity will be systematically reported in a forthcoming study. This is indeed necessary to analyze the performances of QCD focal plane arrays, in term of homogeneity.

\section{conclusion}
In conclusion, we have reported a model of electronic transfers in a multiple quantum well device. Lifetimes associated with all transitions between two subbands are evaluated by considering only interactions between electrons and optical phonons. The resistivity of such a structure around a null bias voltage can then be deduced by Eq~\ref{eq_roa}, which appears as an Einstein relation. An excellent agreement is found between experimental result and our model, which does not include any adjustable parameters.

\begin{acknowledgments}
The authors would like to thank Angela Vasanelli, Aude Leuliet, B\o rge Vinter and Noelle Pottier for usefull dicussion.
\end{acknowledgments}


\begin{thebibliography}{1}
\bibitem{schneider} H.~Schneider and H.C.~Liu, \textit{Quantum well infrared photodetector},(Springer,2006)
\bibitem{laure1} L.~Gendron, M.~Carras, A.~Huynh, V.~Ortiz, C.~Koeniguer and V.~Berger, App.~Phys.~Lett.~{\bf{85}}, 2824 (2004)
\bibitem{graf} M.~Graf, G.~Scalari, D.~Hofstetter, J.~Faist, H.~Beere, E.~Linfield, D.~Ritchie and G.~Davies, App.~Phys.~Lett.~{\bf{84}}, 475, (2004)
\bibitem{laure2} L.~Gendron, C.~Koeniguer, V.~Berger and X.~Marcadet, App.~Phys.~Lett.~{\bf{86}}, 121116 (2005)
\bibitem{brumbastard} J.A.~Brum and G.~Bastard, Phys.~Rev.~B~{\bf{33}}, 1420 (1986) 
\bibitem{brum} J.A.~Brum, T.~Weil, J.~Nagle and B.~Vinter, Phys.~Rev.~B~{\bf{34}}, 2381 (1986) 
\bibitem{rosencher} E.~Rosencher, B.~Vinter, F.~Luc, L.~Thibaudeau, P.~Bois and J.~Nagle, IEEE J.~Quantum~Electron.~{\bf{30}}, 2875 (1994)
\bibitem{liu} H.C.~Liu, App.~Phys.~Lett.~{\bf{60}}, 1507 (1992)
\bibitem{radovanovic} J.~Radovanovi\'c, V.~Milanovi\'c, Z.~Ilkoni\'c, D.~Indjin and P.~Harrison, J.~Appl.~Phys.~{\bf{97}}, 103109 (2005)
\bibitem{aude} A.~Leuliet, A.~Vasanelli, A.~Wade, G.~Fedorov, D.~Smirnov, G.~Bastard and C.~Sirtori, Phys.~Rev.~B~{\bf{73}}, 085311 (2006)
\bibitem{goodnick} S.M.~Goodnick and P.~Lugli, Phys.~Rev.~B~{\bf{37}}, 2578 (1988) 
\bibitem{harrison} P.~Harrison, App.~Phys.~Lett.~{\bf{75}}, 2800 (1999) 
\bibitem{kempa1} K. Kempa, Y. Zhou, J.R. Engelbrecht and P. Bakshi, Phys. Rev. B {\bf{68}}, 085302 (2003) 
\bibitem{kempa2} K.~Kempa et al., Phys.~Rev.~Lett.~{\bf{88}}, 226803 (2002) 
\bibitem{ferreira} R.~Ferreira and G.~Bastard, Phys.~Rev.~B~{\bf{40}}, 1074 (1989) 
\bibitem{rosenchervinter} E.~Rosencher and B.~Vinter, \textit{Optoelectronics}, (Cambridge University Press,2002)
\end{thebibliography}
\end{document}